\documentclass[preprint,aps,superscriptaddress]{revtex4}
\usepackage{graphicx,amssymb,bm,natbib,amsfonts,amsmath}
\usepackage{hyperref}

\begin{document}

\title{First direct observation of Dirac fermions in graphite}

\author{S.Y. Zhou}
\affiliation{Department of Physics, University of California,
Berkeley, CA 94720, USA}
\affiliation{Materials Sciences Division,
Lawrence Berkeley National Laboratory, Berkeley, CA 94720, USA}
\author{G.-H. Gweon}
\affiliation{Department of Physics, University of California,
Berkeley, CA 94720, USA}
\author{J. Graf}
\affiliation{Materials Sciences Division, Lawrence Berkeley National
Laboratory, Berkeley, CA 94720, USA}
\author{A.V. Fedorov}
\affiliation{Advanced Light Source, Lawrence Berkeley National Laboratory, Berkeley, California 94720, USA}
\author{C.D. Spataru}
\affiliation{Department of Physics, University of California,
Berkeley, CA 94720, USA}
\affiliation{Chemical Sciences Division,
Lawrence Berkeley National Laboratory, Berkeley, CA 94720, USA}
\author{R.D.~Diehl}
\affiliation{Department of Physics and Materials Research Institute, Penn State University, University Park, Pennsylvania 16802, USA}
\author{Y. Kopelevich}
\affiliation{Instituto de F\'{i}sica `Gleb Wataghin', Universidade Estadual de Campinas, Unicamp 13083-970, Campinas, Sao Paulo, Brasil}
\author{D.-H. Lee}
\affiliation{Department of Physics, University of California,
Berkeley, CA 94720, USA}
\affiliation{Materials Sciences Division,
Lawrence Berkeley National Laboratory, Berkeley, CA 94720, USA}
\author{Steven G. Louie}
\affiliation{Department of Physics, University of California,
Berkeley, CA 94720, USA}
\affiliation{Materials Sciences Division,
Lawrence Berkeley National Laboratory, Berkeley, CA 94720, USA}
\author{A. Lanzara}
\affiliation{Department of Physics, University of California,
Berkeley, CA 94720, USA}
\affiliation{Materials Sciences Division,
Lawrence Berkeley National Laboratory, Berkeley, CA 94720, USA}

\date{\today}


\maketitle

{\bf Originating from relativistic quantum field theory, Dirac fermions have been recently applied to study various peculiar phenomena in condensed matter physics, including the novel quantum Hall effect in graphene \cite{NovoselovNat,ZhangNat}, magnetic field driven metal-insulator-like transition in graphite \cite{Kopelevich, Xu}, superfluid in $^3$He \cite{He3}, and the exotic pseudogap phase of high temperature superconductors \cite{Franz, WXG2}.  Although Dirac fermions are proposed to play a key role in these systems, so far direct experimental evidence of Dirac fermions has been limited.  Here we report the first direct observation of massless Dirac fermions with linear dispersion near the Brillouin zone (BZ) corner H in graphite, coexisting with quasiparticles with parabolic dispersion near another BZ corner K.  In addition, we report a large electron pocket which we attribute to defect-induced localized states.  Thus, graphite presents a novel system where massless Dirac fermions, quasiparticles with finite effective mass, and defect states all contribute to the low energy electronic dynamics.} 

For most condensed matter systems, the physics is formulated in terms of the nonrelativistic Schr\"odinger equation, and the low energy excitations are quasiparticles with finite effective mass.  For some special systems, e.g. graphene/graphite, where the dispersion is expected to be linear near the Fermi energy E$_F$ and touch E$_F$ only at one point (Dirac point), the physics is described by the relativistic Dirac equation with the speed of light replaced by the Fermi velocity v$_F$.  The low energy excitations in this case are Dirac fermions, which have zero effective mass and a vanishing density of states at the Dirac point.  These massless quasiparticles are proposed to be responsible for various anomalous phenomena observed in these systems \cite{NovoselovNat, ZhangNat, Kopelevich, Gonzalez}.  So far, transport experiments in graphene have shown results in agreement with the presence of Dirac fermions \cite{NovoselovNat,ZhangNat}.  Phase analysis of quantum oscillations in graphite has also suggested coexistence of both Dirac fermions and quasiparticles with finite effective mass \cite{Kopelevich2}.  Here we report the {\it first direct} observation of massless Dirac fermions coexisting with quasiparticles with finite effective mass in graphite, by using angle resolved photoemission spectroscopy (ARPES).  ARPES provides the advantage of directly probing the electronic structure with both energy and momentum information, not accessible by any other measurement.

Figure 1(a) shows an ARPES intensity map measured near the BZ corner H.  The out-of-plane momentum k$_z$ is 0.5 c$^*$, where c$^*$ is the reciprocal lattice constant (see supplemental information for extraction of k$_z$ values).  Following the maximum intensity in this map, a linear $\Lambda$-shaped dispersion can be clearly observed.  The dispersion can be better extracted by following the peak positions in the momentum distribution curves (MDCs), momentum scans at constant energies, shown in panel b.  Here, two peaks in the MDCs disperse linearly and merge near E$_F$.  The Fermi velocity extracted from the dispersion is 0.91$\pm$0.15$\times10^6~m\cdot s^{-1}$, similar to a value 1.1$\times10^6~m\cdot s^{-1}$ reported by a magnetoresistance study of graphene \cite{NovoselovNat}.  We note that at low energy near E$_F$, this linear dispersion is also observed along other cuts through the H point, with similar Fermi velocity.  This linear and isotropic dispersion is in agreement with the behavior of Dirac fermions.

Another way of probing the linear and isotropic dispersion is to study the intensity maps at constant energy.  At E$_F$ (Figure 2(a)), the intensity map shows a small object near H.  The details of this small object will be discussed later.  With increasing binding energy, this object expands and shows a circular shape (panels b-c). We note that only the circular shape in the first BZ is clearly observed.  This is attributed to the dipole matrix element \cite{BZselection}, which suppresses or enhances the intensity in different BZs.  However, taking the three fold symmetry of the sample, this circular shape in the first BZ is expected to extend to other BZs and thus the electronic structure is isotropic near H.  As the energy increases to -0.9 eV, the constant energy map slightly deviates from the circular shape (see arrow in panel d).  This deviation increases with binding energy and a trigonal distortion is clearly observed at -1.2 eV (panel e).  This trigonal distortion is determined by the relevant tight binding parameters for graphite and further studies to analyze this trigonal distortion are in progress.  Overall, Figure 2 shows that from E$_F$ to -0.6 eV, the electronic structure is isotropic in the k$_x$-k$_y$ plane.  Similarly, the Fermi velocity measured within the first BZ is 1.0$\times$10$^6m\cdot s^{-1}$ with a $\le$ 10$\%$ variation along different directions, consistent with the circular constant energy maps shown here.  Combining the results of Figures 1 and 2, we conclude that from E$_F$ to -0.6 eV, the dispersion shows a cone-like behavior near each BZ corner H, similar to what is expected for graphene (panel f).

To resolve the details of the low energy dispersion and the small object at E$_F$ (Figure 2(a)), we show in Figure 3(a) an intensity map measured near H with lower photon energy to give better energy and momentum resolution.  
In the intensity map, one can distinguish two bands dispersing linearly toward E$_F$, as also clear in the MDCs (panel b) where two peaks can be observed for all binding energies.  The extracted dispersion (open circles in panel a) from MDCs (panel b) shows two bands dispersing linearly toward E$_F$, with a minimum separation of 0.020$\pm$0.004 $\AA^{-1}$ at E$_F$.  This linear dispersion near the H point, as well as the isotropic electronic structure shown in Figure 2 from E$_F$ to -0.6 eV, is a basic characteristic of Dirac quasiparticles, which points to the presence of Dirac quasiparticles in the low energy excitations near the H point in graphite.  Furthermore, from the extracted dispersions, the Dirac point is extrapolated to be 50$\pm$20 meV above E$_F$, and thus the small object observed at E$_F$ is a hole pocket.  Assuming an ellipsoidal shape for the hole pocket \cite{Sato}, the hole concentration is estimated to be 3.1$\pm$1.3$\times$10$^{18}$ cm$^{-3}$, from the 0.020 $\AA^{-1}$ separation of the peaks at E$_F$.  This hole concentration is in agreement with reported values \cite{SouleHall, ZhangPRL, Sato}.  The presence of holes with Dirac fermion dispersion is further supported by the angle integrated intensity (panel c), which is proportional to the occupied density of states, barring the matrix element.  In this energy range, a linear behavior, similar to what is expected for Dirac fermions, is observed.  In addition, the energy intersect is at $\approx$ 50 meV above E$_F$, in agreement with the Dirac point energy extrapolated from the dispersions.

Figures 1 to 3 show that near the H point, the low energy excitations in graphite are massless Dirac fermions characterized by linear and isotropic cone-like dispersion, in agreement with transport measurements in graphite where Dirac fermions are suggested to coexist with quasiparticles which have finite effective mass \cite{Kopelevich2}.  To gain direct insight on the different types of quasiparticles, ARPES can provide a unique advantage by directly measuring the effective mass as well as accessing its momentum dependence.  
Figure 4(a) shows the intensity map near another high symmetry point in the BZ corner, the K point.  The dispersion (open circles) shows a parabolic behavior, in sharp contrast to the linear behavior observed near the H point (Figures 1 and 3).  This parabolic dispersion points to the presence of quasiparticles with finite effective mass.  To determine the effective mass, we first extract the low energy dispersion, then fit the MDC and EDC dispersions (panels b, c) with a parabolic function.  In both cases, the effective mass is determined to be 0.069$\pm$0.015 m$_e$, where m$_e$ is the free electron mass.  This effective mass measured by ARPES shows some difference with values reported by transport measurements \cite{ZhangPRL, SouleHall, GaltCR}, 0.052 and 0.038 m$_e$ for electrons and holes respectively. This difference may be due to the fact that transport measurements are not momentum selective, and therefore the mass measured is the average mass over all k$_z$ values.  On the other hand, ARPES is momentum selective and the value for the effective mass is for this specific k$_z$ value only.  Taking this into account, the agreement between these measurements is reasonable.  In summary, the data presented so far show that the low energy excitations in graphite change from massless Dirac fermions with linear dispersion near H (Figures 1-3) to quasiparticles with parabolic dispersion and finite effective mass near K (Figure 4(a)).

We now discuss another interesting feature observed in graphite, i.e., a large electron pocket near E$_F$.  Figure 4(d) shows the intensity map measured in the same experimental conditions as Figure 4(c) except at a different spot.  In panel d, a strong and large electron pocket within 50 meV below E$_F$ is the dominant feature.  Weak signatures of the parabolic $\pi$ band (as in panel c) can still be observed, as the MDC at E$_F$ (panel e) demonstrates.  Here, in addition to the two main peaks (black arrows) corresponding to the electron pocket, a central weak peak (gray arrow) corresponding to the top of the parabolic band can also be distinguished. From the separation ($\approx$ 0.1 $\AA^{-1}$) between the two main peaks at E$_F$, the electron concentration is determined to be 8.0$\pm$0.7$\times$10$^{19}$ cm$^{-3}$, which is an order of magnitude higher than the values reported \cite{ZhangPRL, SouleHall}.  Moreover, from the dispersion (panel d), the effective mass is extracted to be 0.42 $\pm$ 0.07 m$_e$, which is also much larger than any mass reported by transport measurements \cite{ZhangPRL, SouleHall, GaltCR}.  

This large electron pocket, though position dependent, is observed in most of the samples measured, and thus it represents an important feature associated with graphite.  We propose defect-induced localized states as a possible explanation for this large electron pocket, based on the following reasons.  First, the electron concentration and effective mass measured for this electron pocket are much larger than reported values.  Second, although the parabolic $\pi$ band in panels a and c is observed in all the samples measured and in different spots (averaged over $\approx$100 $\mu$m) within the same sample, this large electron pocket strongly depends on the spot position within the same sample.  Third, STM shows that zigzag edges can induce a peak in the local density of states at an energy ($\approx$ -0.03 eV) similar to the electron pocket observed here \cite{STMedges}.  In fact, it has been shown that low concentration of defects (e.g.~edge states, vacancies, {\it etc.}) can induce self-doping to the sample \cite{Antonio, AntonioLoc}.  If this interpretation is correct, then further studies on this large electron pocket may shed light on the magnetic properties of nano-graphite ribbons, since it has been proposed that some defect-induced localized states are magnetic \cite{EdgeStates, nanographite}.

In conclusion, we report the low energy electronic dynamics of graphite, where massless Dirac fermions, quasiparticles with finite effective mass, as well as defect-induced localized states all have important contributions.  These low energy electronic properties are important in understanding the exotic properties not only in graphitic systems, but also in other strongly correlated systems where Dirac fermion dynamics may be important. 

{\bf Supplementary information} 

High resolution ARPES data have been taken at Beamlines 12.0.1 and 7.0.1 of the Advanced Light Source (ALS) in Lawrence Berkeley National Lab with photon energies from 20 to 155 eV  with energy resolution from 15 to 65 meV.  Data were taken from single crystal graphite samples at a temperature of 25 K.  Throughout this paper, the k$_z$ values are estimated using the standard free-electron approximation of the ARPES final state \cite {Himpsel, Hufner, Law}.  The inner potential needed for extracting the k$_z$ value was determined from the periodicity of the detailed dispersion at the Brillouin zone center $\Gamma$A using a wide range of photon energies from 34 to 155 eV\@.  The consistency of the k$_z$ values is confirmed by the degeneracy of the $\pi$ bands near H and a maximum splitting near K (Figure 5),  both in agreement with band structure \cite{Charlier, DresselhausGIC, multilayerG}.

\begin{acknowledgments}
We thank A. Castro Neto, V. Oganesyan, A. Bill, K. McElroy, C.M. Jozwiak, and
D. Garcia for useful discussions; E. Domning and B. Smith for beam line 12.0.1 control
software. This work was supported by the National Science Foundation through Grant
No. DMR03-49361, the Director, Office of Science, Office of Basic Energy Sciences, Division
of Materials Sciences and Engineering of the U.S Department of Energy under Contract
No. DEAC03-76SF00098, and by the Laboratory Directed Research and Development Program
of Lawrence Berkeley National Laboratory under the Department of Energy Contract
No. DE-AC02-05CH11231.
\end{acknowledgments}

\begin{figure}
\includegraphics[width=8.5 cm] {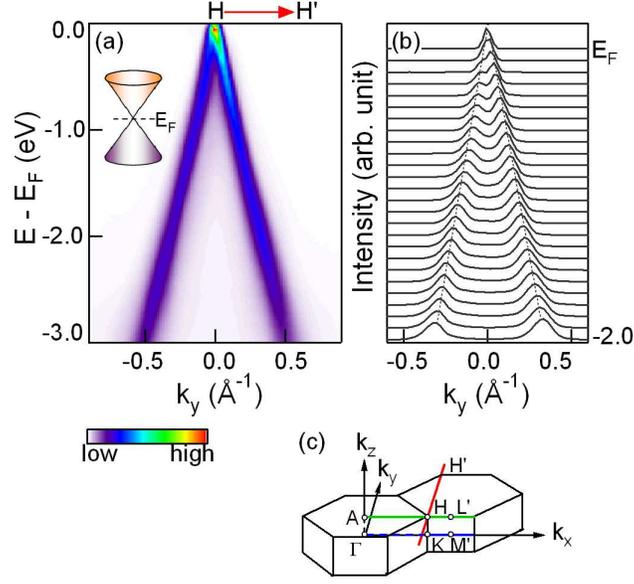}
\label{Figure 1}
\caption{{\bf Linear $\Lambda$-shaped dispersion near the BZ corner H.}  (a) ARPES intensity map taken near the H point (photon energy h$\nu$=140 eV, k$_z\approx0.50~c^*$), along a cut through H and perpendicular to k$_x$ (see red line in the BZ shown in panel c).  The inset shows a schematic diagram of the Dirac cone dispersion near E$_F$ in the three dimensional E-k$_x$-k$_y$ space.  (b) MDCs from E$_F$ to -2.0 eV.  The MDCs are normalized to have the same amplitude and displaced by the same amount so that the dispersion can be directly viewed by following the peak positions at each energy.  The dotted lines are guides to the eyes for the linear-dispersing peaks in the MDCs.  (c) Three dimensional BZ for graphite with high symmetry directions relevant for this paper highlighted with red, green and blue lines.}
\end{figure}

\begin{figure}
\includegraphics [width=12 cm] {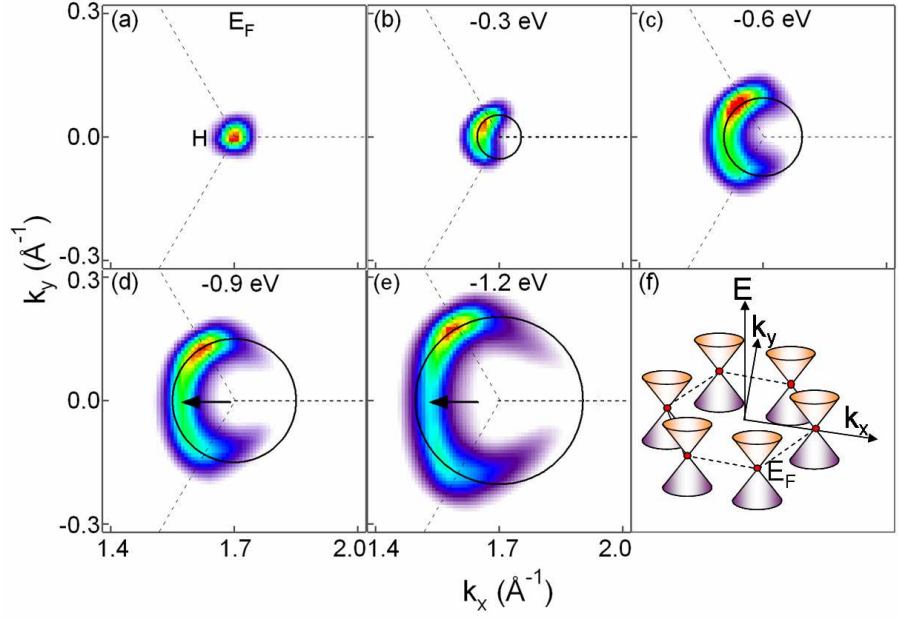}
\label{Figure 2}
\caption{{\bf Constant energy maps taken near the H point, showing that the electronic structure is isotropic in the k$_x$-k$_y$ plane from E$_F$ to -0.6 eV.}  (a-e) ARPES intensity maps near H (h$\nu$=140 eV, k$_z\approx0.50~c^*$) taken at energies from E$_F$ to -1.2 eV.  The circles are guides for the circular intensity pattern near the H point.  Arrows in panels d and e point to deviation from the circle. (f) Schematic diagram of the dispersion for graphene near six BZ corners in the three dimensional E-k$_x$-k$_y$ space.}
\end{figure}

\begin{figure}
\includegraphics[width=10.8 cm]{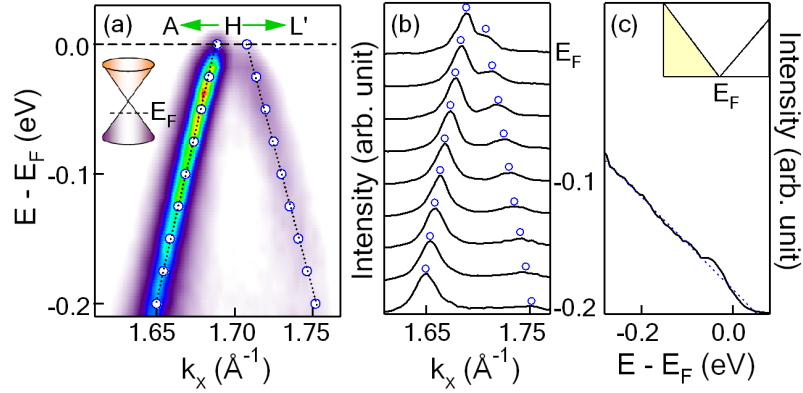}
\label{Figure 3}
\caption{{\bf Detailed low energy dispersion near the H point shows that low energy excitations are Dirac fermions with the Dirac point slightly above E$_F$.}  (a) ARPES intensity map near the H point (h$\nu$=65 eV, k$_z\approx0.45~c^*$) along AHL$^\prime$ direction (green line in the BZ shown in Figure 1(c)).  The open blue circles in panel a show the MDC dispersion. The dotted straight lines are guides for the linear dispersion. (b) MDCs at energies from E$_F$ to -0.2 eV for data shown in panel a.  The blue circles denote the MDC peak positions.  Note that, similar to Figure 2, the intensity of the $\pi$ band is strongly enhanced in the first BZ (H$\rightarrow$A direction), due to the dipole matrix element \cite{BZselection}.  (c) Angle integrated intensity taken near H (h$\nu$=140 eV, k$_z\approx0.50~c^*$).  An overall linear behavior is observed, with some small additional intensity near -0.05 eV.  The origin of this additional intensity is unclear and needs further investigation.  The inset shows the expected density of states for Dirac quasiparticles.  The shaded area shows the occupied density of states.}
\end{figure} 

\begin{figure*}
\includegraphics[width=13.5 cm]{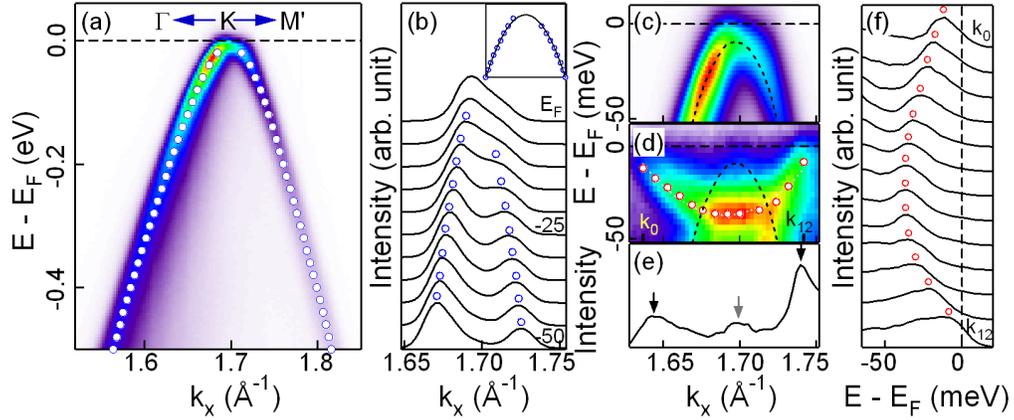}
\label{Figure 4}
\caption{{\bf Detailed dispersion near K, which shows that quasiparticles with finite effective mass and defect-induced localized states also contribute to the low energy electronic dynamics.}  (a) ARPES intensity map near K (h$\nu$=50 eV, k$_z\approx0.08~c^*$) along $\Gamma$KM$^\prime$ direction (blue line in the BZ shown in Figure 1(c)).  The open circles are the dispersions extracted from MDCs.  (b) MDCs from E$_F$ to -50 meV for data in panel a.  The open circles mark the peaks clearly resolved in the data.  The inset shows the MDC dispersion from -10 to -50 meV, with the parabolic fit used to extract the effective mass.  (c) Closer view of data shown in panel a with the energy distribution curve (EDC) dispersion (dotted line).  (d) Intensity map near the K point measured in some parts of the sample, which shows an additional large electron pocket.  The open circles are dispersions extracted from EDCs shown in panel f.  (e) MDC at E$_F$ from data shown in panel d.  The black arrows point to the peaks from the large electron pocket which are separated by $\approx$ 0.1 $\AA^{-1}$, while the gray arrow points to the peak from the $\pi$ band.  (f) EDCs from k$_0$ to k$_{12}$, as indicated in panel d.  Open circles are the peak positions for the large electron pocket.}
\end{figure*}

\begin{figure}
\includegraphics[width=7.4 cm]{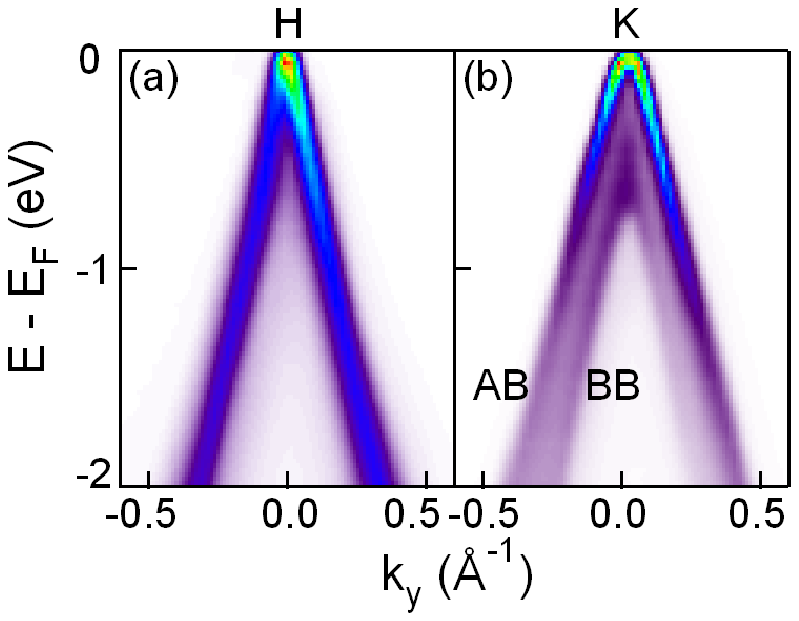}
\label{Figure 5}
\caption{{\bf Dispersions measured near H and K, showing the general consistency of the extracted k$_z$ values.}  (a) Dispersions near H (h$\nu$=140 eV, k$_z\approx0.50~c^*$) along HH$^\prime$ direction, showing that the $\pi$ bands are degenerate.  (b) Dispersions near K (h$\nu$=80 eV, k$_z\approx0.07~c^*$) along direction parallel to HH$^\prime$, where the $\pi$ bands split into bonding (BB) and antibonding (AB) bands.}
\end{figure}


\begin {thebibliography} {99}
\bibitem{ZhangNat} Zhang, Y.B., Tan, Y.-W., Stormer, H. L., and Kim, P. Experimental observation of the quantum Hall effect and Berry's phase in graphene. {\it Nature} {\bf 438}, 201 (2005).
\bibitem{NovoselovNat} Novoselov, K.S., Geim, A.K., Morozov, S.V., Jiang, D., Katsnelson, M.I., Grigorieva, I.V., Dubonos, S.V., and Firsov, A.A.  Two-dimensional gas of massless Dirac fermions in graphene. {\it Nature} {\bf 438}, 197 (2005).
\bibitem{Kopelevich} Kopelevich, Y., Torres, J.H.S., and Slva, R.R. da. Reentrant metallic behavior of graphite in the quantum limit. {\it Phys. Rev. Lett.} {\bf 90}, 156402 (2003).
\bibitem{Xu} Du, X., Tsai, S.-W., Maslov, D.L., and Hebard, R.  Metal-insulator-like behavior in semimetallic bismuth and graphite. {\it Phys. Rev. Lett.}, {\bf 94}, 166601 (2005). 
\bibitem{He3} Volovik, G.E. Field theory in superfluid $^3$He: what are the lessons for particle physics, gravity and high-temperature superconductivity? {\it Proc. Natl. Acad. Sci.} {\bf 96}, 6042 (1999).
\bibitem{WXG2} Rantner, W., Wen, X.G. Electron spectral function and algebraic spin liquid for the normal state of underdoped high T$_c$ superconductors.  {\it Phys. Rev. Lett.} {\bf 86}, 3871 (2001). 
\bibitem{Franz} Franz, M., and Tesanovic, Z.  Algebraic Fermi liquid from phase fluctuations:'topological' Fermions, votex 'Berryons', and QED$_3$ theory of cuprate superconductors.  {\it Phys. Rev. Lett.} {\bf87}, 257003 (2001).  
\bibitem{Gonzalez} Gonz\'{a}lez, J., Guinea, F., and Vozmediano, M.A.H. Unconventional quasiparticle lifetime in graphite. {\it Phys. Rev. Lett.} {\bf 77}, 3589 (1996).
\bibitem{Kopelevich2} Luk'yanchuk, I.A., Kopelevich, Y.  Phase analysis of quantum oscillations in graphite. {\it Phys. Rev. Lett.} {\bf 93}, 166402 (2004).
\bibitem{BZselection} Shirley, Eric L., Terminello, L.J., Santoni, A. and Himpsel, F.J. Brillouin-zone-selection effects in graphite photoelectron angular distributions. {\it Phys. Rev. B} {\bf51}, 13614 (1995).
\bibitem{Sato}Sugawara, K., Sato, T., Souma, S., Takahashi, T., and Suematsu, H. Fermi surface and edge-localized states in graphite studied by high-resolution angle-resolved photoemission spectroscopy.  {\it Phys. Rev. B} {\bf 73}, 045124 (2006).
\bibitem{SouleHall} Soule, D.E. Magnetic field dependence of the Hall effect and magnetoresistance in graphite single crystals.  {\it Phys. Rev.} {\bf 112} 698 (1958).
\bibitem{ZhangPRL} Zhang, Y.B., Joshua P.S., Michael, E.S. A., and Kim, P.  Electri field modulation of galvanomagnetic properties of mesoscopic graphite.  {\it Phys. Rev. Lett.} {\bf 94}, 176803 (2005).
\bibitem{GaltCR} Galt, J.K., Yager, W.A., and Dail, H.W. Jr. Cyclotron resonance effects in graphite.  {\it Phys. Rev.} {\bf 103}, 1586 (1965).
\bibitem{STMedges} Kobayashi, Y., Fukui, K.-I., Enoki, T., Kusakabe, K., and Kaburagi, Y. Observation of zigzag and armchair edges of graphite using scanning tunning microscopy and spectroscopy.  {\it Phys. Rev. B} {\bf 71}, 193406 (2005).
\bibitem{Antonio} Peres, N.M.R., Guinea, F., and Castro Neto, A.H. Electronic properties of disordered two-dimensional carbon. {\it Phys. Rev. B} {\bf 73}, 125411 (2006).
\bibitem{AntonioLoc} Pereira, V.M., Guinea, F., Lopes Dos Santos, J.M.B., Peres, N.M.R., and Castro Neto, A.H.  Disorder induced localized states in graphene.  {\it Phys. Rev. Lett.} {\bf96}, 036801 (2006).
\bibitem{EdgeStates} Nakada, K., Fujita, M., Dresselhaus, G. and Dresselhaus, M.S. Edge state in graphene ribbons: nanometer size effect and edge shape dependence.  {\it Phys. Rev. B} {\bf54}, 17954 (1996).
\bibitem{nanographite} Wakabayashi, K., Fujita, M., Ajiki, H. and Sigrist, M. Electronic and magnetic properties of nanographite ribbons.  {\it Phys. Rev. B} {\bf59}, 8271 (1999).
\bibitem{Law} Law, A.R., Johnson, M. T., Hughes, H.P. Synchrontron-radiation-excited angle-resolved photoemission from single-crystal graphite. {\it Phys. Rev. B} {\bf34}, 4289 (1986).
\bibitem{Himpsel} Himpsel, F.J. Angle-resolved measurements of the photoemission of electrons in the study of solids. {\it Adv.~Phys.}~{\bf32}, 1 (1983).
\bibitem{Hufner} H\"ufner, S. {\em Photoelectron Spectroscopy} (Springer, Berlin, 1995).
\bibitem{Charlier} Charlier, J.-C., Gonze, X., and Michenaud, J.-P. First principles study of the electronic properties of graphite. {\it Phys. Rev. B} {\bf 43}, 4579 (1991).
\bibitem{DresselhausGIC} Dresselhaus, M. S., and Dresselhaus, G. Intercalation compounds of graphite.  {\it Adv. In Phys.} {\bf 51}, 1 (2002).
\bibitem{multilayerG} Nilsson, J., Castro Neto, A.H., Guinea, F., Peres, N.M.R. Electronic properties of graphene multilayers.  Preprint at http://www.arxiv.org/abs/cond-mat/0604106 (2006).
\end {thebibliography}
\end{document}